\documentstyle[pre,aps,epsfig,floats,twocolumn]{revtex}
\begin{document}
\draft
\wideabs {

\title{Spectral theory for the failure of linear control in a nonlinear
stochastic system}

\author{Roman O. Grigoriev, Andreas Handel}

\address{School of Physics, Georgia Institute of Technology,
         Atlanta, GA 30332-0430}

\date{\today} \maketitle

\begin{abstract}

We consider the failure of localized control in a nonlinear spatially extended
system caused by extremely small amounts of noise. It is shown that this
failure occurs as a result of a nonlinear instability. Nonlinear instabilities
can occur in systems described by linearly stable but strongly nonnormal
evolution operators. In spatially extended systems the nonnormality manifests
itself in two different but complementary ways: transient amplification and
spectral focusing of disturbances. We show that temporal and spatial aspects of
the nonnormality and the type of nonlinearity are all crucially important to
understanding and describing the mechanism of nonlinear instability. Presented
results are expected to apply equally to other physical systems where strong
nonnormality is due to the presence of mean flow rather than the action of
control.

\end{abstract}

}

\newpage

It has been known for a long time that systems described by nonnormal evolution
operators (operators with non-orthogonal eigenfunctions) often display rather
surprising dynamics. For instance, turbulence in shear flows often develops for
Reynolds numbers where the basic flow is still {\em linearly} stable. The
critical Reynolds number was found to depend rather sensitively on the geometry
of the system and the roughness of the boundaries. Several studies
\cite{butler,trefethen,gebhardt} have linked the onset of turbulence to a {\em
nonlinear} instability arising from the interaction between the nonlinearity of
the Navier-Stokes equation and the nonnormality of its linearization caused by
significant mean flow. More recently the idea of a nonlinear instability has
been used to explain the disagreement between the predictions of the linear
stability analysis and experimental data for the contact line instability in
gravity driven spreading of thin liquid films \cite{bertozzi}. Nonnormality can
also arise in the absence of mean flow as a result of localized feedback
control \cite{egolf,nonnormal}. In this paper we use the idea of a nonlinear
instability to explain the failure of localized control of a spatially extended
nonlinear system in the presence of extremely weak noise. We extend and refine
ideas described in \cite{trefethen,egolf} by incorporating the information
about the spatial degrees of freedom.

All studies of strongly nonnormal systems conducted up to now have analyzed the
mechanism for nonlinear instability by concentrating only on the {\em temporal}
dynamics. Although the importance of {\em spatial} degrees of freedom is
generally recognized, the complexity of the problem usually prevents consistent
spatiotemporal analysis. As the subsequent discussion shows, spatial
information is crucial for understanding the mechanism that leads to nonlinear
instability, which involves transient amplification of deviations produced by
nonlinear terms. However, before developing the spatiotemporal description, it
will be useful to review some results of the temporal analysis. Following
\cite{trefethen} let us consider a system
\begin{equation}
\label{eq_toy}
\dot\phi=L\phi+f(\phi),
\end{equation}
where $L$ is a stable linear operator and $f$ is a nonlinear function of its
argument. In a purely linear system the disturbance decays asymptotically in
time. However, if $L$ is nonnormal, this exponential decay can be preceded by a
transient. The strength of nonnormality can be determined by the transient
amplification factor 
\begin{equation}
\label{eq_gamma}
\gamma\equiv\max_{t,\phi(0)}\frac{\|\phi(t)\|}{\|\phi(0)\|}
=\max_t\left\|e^{Lt}\right\|,
\end{equation}
such that $\gamma=1$ for $L$ normal. The maximal transient amplification
is achieved for the optimal initial disturbance $\phi(0)=\phi_{opt}$ at
the optimal time $t=t_{opt}$ \cite{farrell}.

Any initial disturbance $\phi(0)$ with a nonvanishing component along
$\phi_{opt}$ will be transiently amplified as well. For small disturbances,
$\sigma=\|\phi(0)\|\ll 1$, the linear terms will dominate, so at the peak of
the transient we will have $\|L\phi(t_{opt})\|\sim \|\phi(t_{opt})\|\sim
\gamma\sigma$. For a quadratic nonlinearity, $\|f(\phi(t_{opt}))\|\sim
(\gamma\sigma)^2$, so it will produce an integrated deviation in (\ref{eq_toy})
of order $t_{opt}(\gamma\sigma)^2$ in the same amount of time it takes the
linear operator to amplify the initial disturbance by $O(\gamma)$. The temporal
analysis makes an implicit assumption that generically this deviation has again
a nonvanishing component along the optimal disturbance $\phi_{opt}$, so it will
be transiently amplified by $L$ in the same way as the initial disturbance. (As
we will see below, this assumption can break down for spatially extended
systems due to their high symmetry.) The deviation due to nonlinear terms will
grow producing a positive feedback loop, if $t_{opt}(\gamma\sigma)^2\gtrsim
\sigma$, and decay otherwise. The critical magnitude of a disturbance needed to
bootstrap the nonlinear instability will, therefore, scale like $\sigma\sim
t_{opt}^{-1}\gamma^{-2}$ for a quadratic nonlinearity. While in some systems
like channel flow $t_{opt}\sim\gamma$, more often the dependence on $t_{opt}$
is too weak to be of any importance, e.g., for both coupled map lattices
\cite{egolf,cml} and partial differential equations \cite{nonnormal} with
localized control $t_{opt}\sim\log\gamma$. In the latter case we have a simple
power law scaling $\sigma\sim\gamma^\alpha$ with an exponent $\alpha=-2$.

\begin{figure}[t]
\centering
\mbox{\epsfig{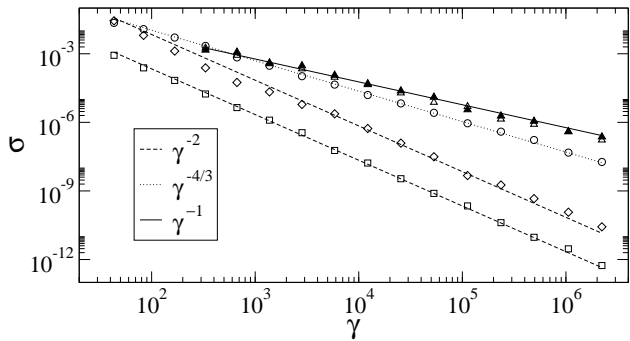}}
\vskip 3mm
\caption{The noise level for which localized linear control fails as a function
of the transient amplification factor for different types of nonlinearities
($\phi^2$ - squares, $\phi_x\phi$ - diamonds, $\phi^3$ - open triangles,
$\phi^4$ - circles, $\phi^5$ - filled triangles). Straight lines are
theoretical fits with slopes given by (\ref{eq_exponent}).}
\label{fig_noise}
\end{figure}

Sometimes, temporal analysis is sufficient and does give correct predictions
for spatially extended systems. For instance, a controlled lattice of coupled
quadratic maps \cite{egolf} has indeed produced the scaling exponent
$\alpha=-2$. However, sometimes the predictions of the temporal analysis are
clearly wrong, suggesting that the spatial structure of disturbances plays an
important role and cannot generally be ignored. A particularly simple example
of the failure of temporal analysis is provided by the generalized
Ginzburg-Landau equation (GGLE)
\begin{equation}
\label{eq_gl}
{\phi}_t=\phi+\phi_{xx}+f(\phi)+\xi,
\end{equation}
which (aside from the stochastic term $\xi$) is of the same form as
(\ref{eq_toy}). The dynamics of GGLE can be made linearly stable via feedback
control imposed at the boundaries
\begin{equation}
\label{eq_bc}
{\phi}(0,t)=0,\qquad \phi'(L,t)=\int_0^lK(x)\phi(x,t)dx,
\end{equation}
where $K(x)$ is an appropriately chosen gain function. As we have shown
previously \cite{nonnormal}, the application of spatially localized control
(\ref{eq_bc}) to the spatially extended system (\ref{eq_gl}) makes the
linearized dynamics strongly nonnormal. We therefore expect the nonlinear
instability to play a prominent role in destabilization as large transient
amplification makes the dynamics extremely susceptible to noise. Numerical
simulations of (\ref{eq_gl}) with a power law nonlinearity $f(\phi)\propto
\phi^n$ and random noise $\xi$ uniformly distributed on $(-\sigma,\sigma)$ show
(see Fig.~\ref{fig_noise}) that the critical noise level resulting in the
failure of linear control scales as a power law $\sigma\sim\gamma^\alpha$, with
an exponent well approximated by
\begin{equation}
\label{eq_exponent}
\alpha=\left\{\matrix{
-n/(n-1), & n=2,4,\cdots,\cr
-1,\hfill & n=3,5,\cdots.}\right.
\end{equation}
The exponent for $n$ even is correctly predicted by a properly generalized
temporal analysis. Indeed, for controlled GGLE $t_{opt}\sim\log\gamma\sim l$,
so comparing stochastic disturbances of order $\sigma$ with the distortions of
order $(\gamma\sigma)^n$ produced by a combination of transient growth and
nonlinearity, we immediately obtain the scaling $\sigma\sim \gamma^{-n/(n-1)}$.
However, the corresponding exponent is inconsistent with our numerical results
for $n$ odd.

\begin{figure}[t]
\centering
\mbox{\epsfig{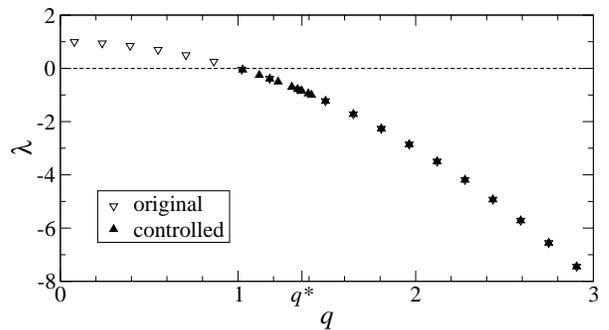}}
\vskip 3mm
\caption{The wavenumbers $q_n$ and the corresponding eigenvalues
$\lambda_n=1-q_n^2$ for the original and controlled system. The system size
here and in the rest of the paper is $l=20$.}
\label{fig_spect}
\end{figure}

This discrepancy calls for the development of a more accurate theory that will
be capable of explaining the effects of arbitrary types of nonlinearities. In
particular, we would like to know why an advective term $\phi_x\phi$ produces
the same scaling as a simple quadratic nonlinearity $\phi^2$ despite their
different symmetry properties. As it turns out, the explanation can be obtained
rather easily by conducting a spatiotemporal analysis of the bootstrapping
mechanism. Indeed, transient amplification represents just one aspect of the
nonnormal dynamics. The other aspect ignored by the temporal analysis is the
focusing of the initial disturbances in the direction of the most strongly
nonnormal eigenfunctions.

It is easy to see that due to the translational invariance of the linear
operator $L=1+\partial_x^2$, its eigenfunctions are sinusoidal, with or without
control. The boundary conditions (\ref{eq_bc}) determine the wavenumber $q$ of
an eigenfunction and the corresponding eigenvalue $\lambda=1-q^2$, such that
the eigenfunction is stable when $q>1$ and unstable otherwise. In particular,
the eigenfunctions of the original system (no feedback, $K(x)= 0$) are 
$u_k(x)=\sin(q_kx)$ with $q_k=\pi(k-1/2)/l$, so at least one will be unstable
for $l>\pi/2$. All eigenfunctions $v_k(x)=\sin(q'_kx)$ of the controlled system
are stable with wavenumbers $q'_k>1$. In general, $q'_k$ might be complex, but
we can always force them to be real. This is done throughout the paper by
calculating $K(x)$ using Linear-Quadratic control \cite{dorato}. Appearance of
complex eigenvalues does not affect the following analysis. As
Fig.~\ref{fig_spect} shows, feedback (\ref{eq_bc}) shifts all wavenumbers from
the unstable band $Q_{+}=(0,1)$ into the stable band $Q_{-}=(1,\infty)$. The
new wavenumbers $q'_k$ cluster most tightly in a rather narrow band
$Q_\parallel$ centered at  $q^*\approx 1.36$. As the size $l$ of the system
grows, an increasing number of eigenfunctions of the original system becomes
unstable and gets squeezed into $Q_\parallel$ by feedback. As a result the
distance between wavenumbers in $Q_\parallel$ shrinks and the corresponding
eigenfunctions become increasingly aligned (nonnormal) \cite{nonnormal}.

\begin{figure}[t]
\centering
\mbox{\epsfig{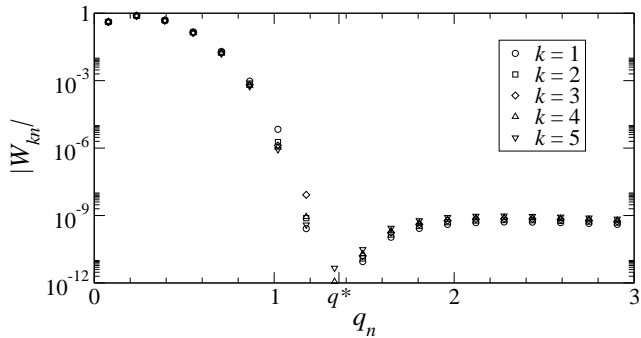}}
\vskip 3mm
\caption{Fourier spectra of the first five adjoint eigenfunctions of the
controlled system. The expansion is in the basis $\{u_n\}$ of eigenfunctions of
the original system, $w_k(x)=\sum_n W_{kn}\sin(q_nx)$.}
\label{fig_adjnt}
\end{figure}

\begin{figure}[t]
\centering
\mbox{\epsfig{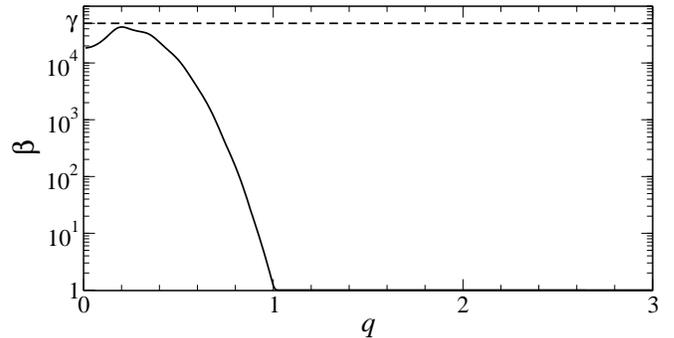}}
\vskip 3mm
\caption{Transient amplification factor for sinusoidal initial disturbances as
a function of their wavenumber.}
\label{fig_trans}
\end{figure}

Now, let us consider what happens with an initial disturbance $\phi(x,0)$. Let
us first concentrate on the linear effects. As the set $\{v_k\}$ is complete,
in the absence of noise the general solution of the linearized equation
(\ref{eq_gl}) is given by
\begin{equation}
\label{eq_rlin}
\phi(x,t)=\sum_{k=1}^\infty c_k v_k(x)\exp(\lambda'_kt),
\end{equation}
where $\lambda'_k=1-(q'_k)^2$. The coefficients $c_k$  can be found using the
adjoint eigenfunctions $w_k(x)$: 
\begin{equation}
\label{eq_coeff}
c_k=\frac{(w_k(x),\phi(x,0))}{(w_k(x),v_k(x))},
\end{equation}
where we assume that the eigenfunctions are normalized such that
$\|w_k\|=\|v_k\|=1$. As the eigenvalues $\lambda'_k$ are all stable, it is
clear that transient amplification can only result from large values of the
coefficients $c_k$. For $c_k$ to become large two conditions have to be
satisfied. First, the numerator in (\ref{eq_coeff}) should not be small, i.e.,
the initial disturbance should not be orthogonal to the adjoint eigenvector
$w_k$. As adjoint eigenvectors satisfy the orthogonality condition
$(w_k,v_m)=0$ for $k\ne m$, their Fourier spectra are localized to the unstable
band $Q_{+}$ (see Fig. \ref{fig_adjnt}). Therefore, only disturbances with
significant spectral content in $Q_{+}$ will be transiently amplified. (Such
disturbances will grow indefinitely in the uncontrolled system; control makes
this growth transient.) This is illustrated in Fig.~\ref{fig_trans} which shows
the transient amplification for sinusoidal initial disturbances:
\begin{equation}
\beta(q)=\max_t\left.\frac{\|\phi(x,t)\|}{\|\phi(x,0)\|}
\right|_{\phi(x,0)=\sin(qx)}.
\end{equation}

\begin{figure}[t]
\centering
\mbox{\epsfig{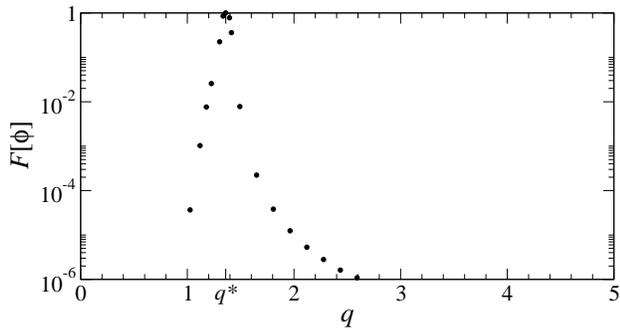}}
\vskip 3mm
\caption{Normalized Fourier spectrum for a linear system driven by random noise
uniformly distributed on $(-\sigma,\sigma)$.}
\label{fig_linear}
\end{figure}

The second condition is that the denominator in (\ref{eq_coeff}) should be
small, which can only happen for strongly nonnormal eigenfunctions $v_k$ with
$q'_k\in Q_\parallel$. As a result, the Fourier spectrum of the transiently
amplified disturbances will be strongly focused into the band $Q_\parallel$.
This focusing effect can be clearly seen in Fig.~\ref{fig_linear} which shows
the spectrum of the linearized GGLE driven by random noise. The spectrum is
computed for the ``worst case perturbation,'' as this is the type of
disturbance that leads to both the failure of linear control and more generally
to the onset of nonlinear instability. Specifically, the state is expanded in
the basis $\{v_k\}$
\begin{equation}
\label{eq_stoch}
\phi(x,t)=\sum_{k=1}^\infty c_k(t)\sin(q'_kx)
\end{equation}
and the "worst case" spectrum $F_k[\phi]$ is obtained by finding the maximal
values of $|c_k(t)|$ for each $k$. The Fourier coefficients outside of
$Q_\parallel$ are seen to be exponentially small. These two intimately related
aspects of the nonnormal dynamics -- transient amplification and focusing --
are likely to be quite common in other strongly nonnormal spatially extended
systems. For instance, a similar clustering of eigenfunctions is found in a
model describing thermally driven spreading of liquid films \cite{thermal}.

Having understood the linear dynamics, let us now consider what happens when
nonlinearities come into play. For sufficiently small $\sigma$ the
nonlinearities will hardly change the linear dynamics. Their effect will be
limited to ``filtering'' the transiently amplified disturbances, changing their
magnitude and spectral content. As the transiently amplified disturbances have
a very narrow spectrum, the spectrum of the signal produced by the nonlinear
terms will also consist of several narrow peaks, as long as we consider
nonlinearities of the power law type $f(\phi)=\phi^n$ with moderate $n$. (High
powers are not interesting as the scaling exponents (\ref{eq_exponent}) for the
even and odd powers become indistinguishable. Besides, most physically
interesting nonlinearities have low powers.) 

\begin{figure}[t]
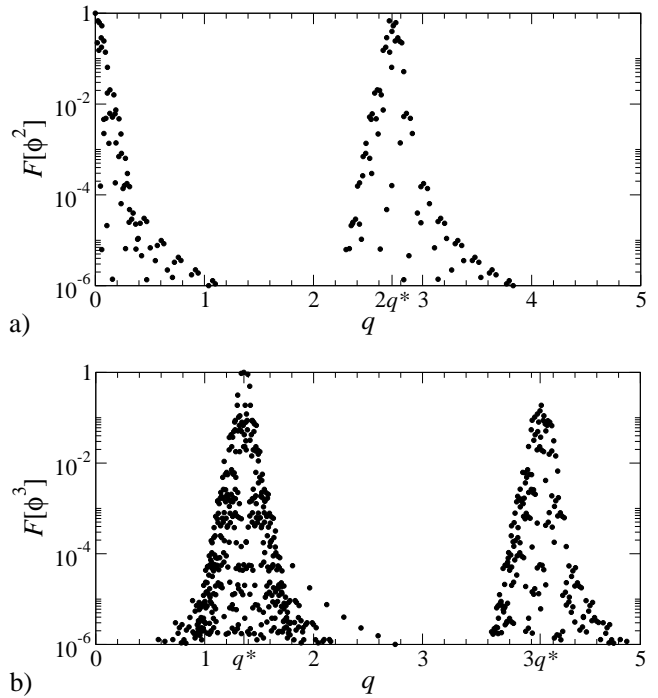

\centering
\mbox{\epsfig{figure=quadr.eps,height=1.75in}}
\vskip 3mm
\mbox{\epsfig{figure=cubic.eps,height=1.75in}}
\vskip 3mm
\caption{Normalized Fourier spectrum of (a) the quadratic term, $\phi^2$, and
(b) the cubic term, $\phi^3$.}
\label{fig_nonlin}
\end{figure}

For instance, the spectrum of a quadratic nonlinearity, be it $\phi^2$ or
$\phi_x\phi$, will only contain frequencies which are either sums or
differences of frequencies $q'_k$, i.e., 0, $|q'_m-q'_n|$, $2q'_m$, and
$q'_m+q'_n$. As the Fourier coefficient corresponding to the frequency
$|q'_m\pm q'_n|$ is of order $c_kc_m$, the only significant contributions are
produced when both $q'_m$ and $q'_n$ lie in $Q_\parallel$. This means that the
spectrum of the quadratic term will be localized near $q=0$ and $q=2q^*$ (see
Fig.~\ref{fig_nonlin}a). The disturbances with $q\approx 2q^*$ are strongly
damped, so the primary effect of most any quadratic nonlinearity will be to
transfer the excitations from the band $Q_\parallel$ back into the band $Q_+$,
where they will again be transiently amplified. After one full cycle involving
transient amplification, focusing, and nonlinear filtering, an initial (e.g.,
stochastic) disturbance of order $\sigma$ will produce a deviation of order
$(\gamma\sigma)^2$. Therefore, for $(\gamma\sigma)^{2}\gtrsim\sigma$ the low
wavenumber disturbances will be driven predominantly by the nonlinear term,
bootstrapping a nonlinear instability. A similar picture will be observed for
$f(\phi)=\phi^n$ with $n=4,6,\cdots$. The spectrum of $f(\phi)$ will contain a
strong component in $Q_+$ and the bootstrapping will occur for
$(\gamma\sigma)^n\gtrsim\sigma$, so the critical noise will scale like
$\gamma^\alpha$, with $\alpha$ given by (\ref{eq_exponent}).

The case of odd powers is substantially different. For instance, the spectrum
of a cubic nonlinearity, $f(\phi)=\phi^3$, will only contain wavenumbers
$|q'_m\pm q'_n\pm q'_k|$. As the corresponding Fourier coefficients will be of
order $c_mc_nc_k$, the spectrum will be strongly localized near $q=q^*$ and
$q=3q^*$. Furthermore, as Fig.~\ref{fig_nonlin} shows, the spectral peaks of
the nonlinear terms {\em do not} broaden. Therefore, a cubic nonlinearity will
not transfer excitations from $Q_\parallel$ back to $Q_+$, and no bootstrapping
will occur. The quintic nonlinearity, $f(\phi)=\phi^5$, is expected to produce
similar results as its spectrum will be localized near $q^*$, $3q^*$, and
$5q^*$, and so on. Destabilization will nevertheless occur for any power $n$
when the nonlinear terms become of the same order of magnitude as the linear
terms, $(\gamma\sigma)^n\sim \gamma\sigma$, i.e., when $\gamma\sigma=O(1)$. At
this point, the predictions of linear stability analysis become invalid.
Therefore, for $n$ odd the critical noise will scale like $\gamma^{-1}$,
justifying the second part of (\ref{eq_exponent}). The result for even powers
is not changed, since  nonlinear instability occurs for levels of noise much
smaller than those at which linear stability analysis breaks down.

Summing up, we can conclude that, at least for a simple equation such as the
stochastically driven GGLE studied here, the failure of localized control can be
explained by a straightforward spectral analysis of transient dynamics.
Moreover, spatiotemporal analysis appears to be crucial for understanding the
mechanism of nonlinear instabilities in spatially extended systems in general.
In particular, as the focusing effect described in this paper is an inherent
feature of strongly nonnormal dynamics, its applicability is not constrained to
the control problem considered here. A similar analysis could provide valuable
insights into stability and control of shear flows, driven contact lines, and
magnetic plasmas \cite{camargo}.



\begin{references}


\bibitem{butler} K. M. Butler and B. F. Farrell,
Phys. Fluids A {\bf 4}, 1637 (1992)

\bibitem{trefethen} L. N. Trefethen, A. E. Trefethen, S. C. Reddy, and T. A.
Driscoll, Science {\bf 261}, 578 (1993).

\bibitem{gebhardt} T. Gebhardt and S. Grossmann,
Phys. Rev. E {\bf 50}, 3705 (1994).



\bibitem{bertozzi} A. L. Bertozzi and M. P. Brenner,
Phys. Fluids {\bf 9}, 530 (1997);
L. Kondic and A. L. Bertozzi,
Phys. Fluids {\bf 11}, 3560 (1999).

\bibitem{egolf} D. A. Egolf and J. E. S. Socolar, 
Phys. Rev. E {\bf 57}, 5271 (1998).

\bibitem{nonnormal} R. O. Grigoriev and A. Handel, 
to be published by Phys. Rev. E, arxiv.org/abs/nlin.PS/0207023.

\bibitem{farrell} B. F. Farrell and P. J. Ioannou,
J. Atmospheric Sci. {\bf 53}, 2025 (1996)

\bibitem{cml} R. O. Grigoriev, M. C. Cross, H. G. Schuster,
Phys. Rev. Lett. {\bf 79}, 2795 (1997).

\bibitem{dorato} P. Dorato, C. Abdallah and V. Cerrone, {\it Linear-Quadratic
Control} (Prentice Hall, New Jersey, 1995)

\bibitem{thermal} R. O. Grigoriev, submitted to Phys. Fluids,
arxiv.org/abs/nlin.PS/0207024

\bibitem{camargo} S. J. Camargo, M. K. Tippett, and I. L. Caldas,
Phys. Plasma {\bf 7}, 2849 (2000).




\end{references}
\end{document}